\documentclass[12pt]{article}

\usepackage{amssymb}
\usepackage{amsmath}
\usepackage{mathrsfs}
\usepackage{bm}
\usepackage{graphicx}
\usepackage{hyperref}
\begin{document}
\begin{titlepage}
%\begin{flushright}
%v2.\today
%\end{flushright}
\begin{center}
\begin{large}
{\bf Unconstrained Variables and Equivalence Relations\\ for\\ Lattice Gauge Theories}
\end{large}

\vskip1truecm

Stam Nicolis\footnote{\tt E-Mail: Stam.Nicolis@lmpt.univ-tours.fr}

\vskip1truecm
{\sl 
CNRS--Laboratoire de Math\'ematiques et Physique Th\'eorique (UMR 6083)\\
F\'ed\'eration ``Denis Poisson'' (FR 2964)\\
Universit\'e ``Fran\c{c}ois Rabelais'' de Tours\\
Parc Grandmont, 37200 Tours, France
}
\end{center}

\vskip1truecm

\begin{abstract}
We write the partition function for a lattice gauge theory, with compact gauge
group, exactly in terms of
unconstrained variables  and show that, in the mean field approximation, 
the dynamics of pure gauge theories, invariant under compact, continuous,
groups of rank 1 is the same for all.  We explicitly obtain the equivalence
 for the case of $SU(2)$ and $U(1)$ and show that it obtains, also, 
if we consider saddle point configurations that are not,
 necessarily, uniform, but only proportional to the identity for both groups.  
This implies that the phase diagrams of the (an)isotropic $SU(2)$ 
theory and the (an)isotropic $U(1)$ theory in any dimension are identical, 
within this approximation, up to a re-evaluation of the numerical values of the 
coupling constants at the transitions. 
Only nonuniform field configurations, that, also, belong to 
higher dimensional representations for Yang--Mills fields, will be able to  probe the difference between them. 

We also show under what conditions the global symmetry of an  anisotropic
term in the lattice action can be promoted to a gauge symmetry of the theory 
on layers and point out how deconstruction and flux compactification scenaria
may thus be studied on the lattice. 

\end{abstract}
\end{titlepage}
The effective potential is a standard tool for obtaining the phase diagram of
a field theory in the continuum~\cite{coleman_weinberg}. 
Its computation for lattice gauge theories is complicated by the fact that the 
link variables are constrained: they take values in the group, which is
typically compact when gravity is not taken into account.
This is helpful for numerical work, since it is easier to sample a 
compact space than a non-compact one (however the fact that the constraint is
local does render the sampling difficult and slow and is an obstacle to
developing cluster algorithms for gauge theories). 

It is therefore useful to see whether and how
 the constraints can indeed be solved in
a way that is helpful both for numerical and for analytical calculations. The
idea is not new~\cite{brezin_drouffe,drouffe_zuber}, but has been associated
too heavily with the mean field approximation to lattice gauge theories. 
 One purpose of the present note is to try to 
eliminate this misunderstanding and to 
show that the trick leads to an expression that enjoys all the symmetries of 
the original theory. We will use it, in particular, to elucidate the
symmetries of theories, whose fields live on the links of the lattice and try
to understand when these symmetries can (or cannot) be consistently gauged.

For an abelian gauge group the idea is realized as follows: Insert in the 
partition function 
\begin{equation}
\label{ZJ}
Z[J] = \int [{\mathscr D} U] e^{-S[U]+a^D\sum_n (J^\dagger\cdot U + J\cdot
  U^\dagger)}\equiv e^{-W[J]}
\end{equation}  
the following expression~\cite{brezin_drouffe,drouffe_zuber} 
\begin{equation}
\label{unity}
1 = \int\prod_{\mathrm{links}} dV_l^R dV_l^I
\delta(\mathrm{Re}(U_\mu(n))-V_l^R)\delta(\mathrm{Im}(U_\mu(n))-V_l^I) 
\end{equation}
and use the integral representation of the delta functions
\begin{equation}
\delta(\mathrm{Re}(U_\mu(n))-V_l^R) =
\int_{-\infty}^\infty\frac{d\alpha_l^R}{2\pi} e^{\mathrm{i}\alpha_l^R(\mathrm{Re}(U_\mu(n))-V_l^R)}
\end{equation}
(and similarly for the imaginary part). The partition function  takes the
form
\begin{equation}
\label{Zunc}
Z[J] = \int\left[\prod_\mathrm{links} dV_l^R dV_l^I
\frac{d\alpha_l^R}{2\pi}\frac{d\alpha_l^I}{2\pi}\right]
e^{-S[V_l^R,V_l^I]-\mathrm{i}\sum_l(\alpha_l^R V_l^R + \alpha_l^I V_l^I) +\sum_l(J_l^R
  V_l^R + J_l^I V_l^I)} \int{\mathscr D} U e^{\mathrm{i}(\alpha_l^R U_l^R + \alpha_l^I U_l^I)}
\end{equation}
The action, $S[V_l^R, V_l^I]$, becomes a function of unconstrained link
variables, $V_l^R$ and $V_l^I$.  
We recognize the 1--link integral over the gauge group~\cite{brezin_gross,broweretal}
\begin{equation}
\label{1linkintegral}
e^{w(\alpha_l^R,\alpha_l^I)}\equiv \int{\mathscr D} U 
e^{\mathrm{i}(\alpha_l^R U_l^R+\alpha_l^I U_l^I)}
\end{equation}
and we can identify the effective action,$S_\mathrm{eff}$,
 over the {\em unconstrained}
variables, $(\alpha_l^R,\alpha_l^I, V_l^R, V_l^I)$,
\begin{equation}
\label{effaction}
S_\mathrm{eff}[\alpha_l^R,\alpha_l^I, V_l^R, V_l^I]\equiv 
S[V_l^R, V_l^I] +\mathrm{i}\sum_l\left(\alpha_l^R V_l^R + \alpha_l^I
V_l^I\right) -\sum_l w(\alpha_l^R, \alpha_l^I)
\end{equation}
For a non-abelian group, where the link variables are unitary matrices, matrix
model techniques~\cite{brezin_gross,broweretal} lead to the same
expression--the trace over the appropriate representation is implicit. 
The real and imaginary parts are replaced by hermitian and anti-hermitian
parts respectively.
 
In this expression the group integral, $w(\alpha_l^R, \alpha_l^I)$ and the 
``constraint term, $\alpha_l^R V_l^R + \alpha_l^I V_l^I$, are both local and
their contribution to the action factorizes over the links exactly. Were the
Wilson action also factorizable, the single-link approximation would be exact. 

To ensure that we have indeed solved the constraints, we must check how the
symmetries are realized in terms of the unconstrained variables. In terms of the
constrained variables, $U_\mu(n)$, the action is invariant under the
transformations $U_\mu(n)\to v_n U_\mu(n) v_{n+\mu}^\dagger$, where $v_n$ and 
$v_{n+\mu}$ are group elements that live at the sites $n$ and $n+\mu$
respectively, the ends of the link between them  
(and no summation over the repeated $\mu$ index is carried out). 

For the $U(1)$ case it is useful to rewrite the effective action in terms of 
the combinations
$V_l^{\pm}\equiv V_l^R\pm \mathrm{i} V_l^I$ and $\alpha_l^{\pm}\equiv 
\alpha_l^R\pm\mathrm{i}\alpha_l^I$:
$$
S_\mathrm{eff}[V_l^\pm,\alpha_l^\pm] = S[V_l^+,V_l^-] +
\frac{\mathrm{i}}{2}\sum_l\left(\alpha_l^+V_l^-+\alpha_l^- V_l^+\right) -
\sum_l w(\alpha_l^+,\alpha_l^-)
$$
This expression is quite remarkable. The lattice action of the original gauge
theory has been rewritten in terms of four fields, $V_l^\pm$ and
$\alpha_l^\pm$,  that live on each link. The $\alpha_l^\pm$ do not propagate,
since their action factorizes over the links--they enforce the
constraints. The $V_l^\pm$ do propagate, since $S[V_l^+, V_l^-]$ does not 
factorize over the links. Thus we have achieved a consistent separation of the
propagating from the non-propagating degrees of freedom of the gauge field,
without having to introduce a gauge-fixing condition. This is possible because
the group is compact.

Furthermore these
fields are charged under the gauge group:
 The action is invariant under the--local--transformations
$$
\begin{array}{cccc}
\displaystyle
\delta V_l^+ = -\mathrm{i}\theta_l V_l^+ &
\displaystyle
\delta V_l^- = \mathrm{i}\theta_l V_l^- &
\displaystyle
\delta \alpha_l^+ = -\mathrm{i}\theta_l \alpha_l^+ &
\delta \alpha_l^- = \mathrm{i}\theta_l \alpha_l^- \\
\end{array}
$$
which are those of fields, living on the links and 
carrying charges $\pm 1$ under the gauge group $U(1)$. These charges may be
identified with the helicity states of the gauge field. 
It is also easy to check that these transformations leave the measure 
invariant also. Therefore the change of variables from $U_\mu(n)$ to
$V_l^\pm$, $\alpha_l^\pm$ does not change the physics. In fact the expression 
$S[V_l^+,V_l^-]+w(\alpha_l^+, \alpha_l^-)$ is invariant under independent, 
local, $U(1)$ transformations for $V$ and $\alpha$--i.e. it is invariant under $U(1)\times U(1)$; the ``mixing term'', 
$\alpha_l^+ V_l^-+\alpha_l^- V_l^+$ is invariant only under the diagonal
$U(1)$ subgroup, which is, thus, the symmetry group of the full expression, as
expected. 

 The effective action appears to be complex. This is, however, not an 
unavoidable conclusion. The theory of Fourier transforms teaches us that the 
Fourier transform of a function that is reflection--positive 
is a real function. The passage from the constrained variables to the
unconstrained ones is a Fourier transform and the Wilson action, 
we consider here, is, indeed, reflection-positive~\cite{menotti_pelissetto}, 
thus the partition function is real and we may find a contour that renders this
explicit~\cite{drouffe_zuber}: For instance, a Wick rotation in the $\alpha_l$
integrals, $\alpha_l^R = -\mathrm{i}\widehat{\alpha}_l^R, 
\alpha_l^I = -\mathrm{i}\widehat{\alpha}_l^I$ (for the case of $U(1)$) 
The effective action then takes
the form 
\begin{equation}
\label{effactionE}
S_\mathrm{eff}[\widehat{\alpha}_l^R,\widehat{\alpha}_l^I, V_l^R, V_l^I]\equiv 
S[V_l^R, V_l^I] +\sum_l\left(\widehat{\alpha}_l^R V_l^R + 
\widehat{\alpha}_l^I
V_l^I\right) -\sum_l w(\widehat{\alpha}_l^R, \widehat{\alpha}_l^I) -
\sum_l\left(J_l^R V_l^R + J_l^I V_l^I\right)
\end{equation}
and the partition function is given by 
\begin{equation}
\label{ZuncE}
Z[J] = 
\int_{-\infty}^{\infty}\left[\prod_\mathrm{links} dV_l^R dV_l^I\right]
\int_{-\mathrm{i}\infty}^{\mathrm{i}\infty}
\left[\frac{d\widehat{\alpha}_l^R}{2\pi\mathrm{i}}
\frac{d\widehat{\alpha}_l^I}{2\pi\mathrm{i}}\right]
e^{-S_\mathrm{eff}[V_l^R, V_l^I, \widehat{\alpha}_l^R,
    \widehat{\alpha}_l^I,J_l^R, J_l^I]}
\end{equation}
 If the lattice action is not reflection positive, or is even complex
 itself, then it is an interesting question, whether this representation is
 more, equally or less useful than the original one in terms of the
 constrained variables(i.e. the group valued gauge links). In our case the
 advantages are clearly apparent. 

Let us now try the same procedure for the gauge group $SU(2)$. In this case
the link variables are $U_\mu(n) = \exp(\mathrm{i}\bm{u}_\mu(n)\cdot\bm{T})$, where 
the vector $\bm{u}$ has real entries and $\bm{T}$ are the generators of $SU(2)$
in the appropriate representation that satisfy the commutation relations
$[T^A, T^B]=\mathrm{i}\varepsilon^{ABC}T^C$. Of particular interest are the
cases where the link variables belong to the fundamental representation ($T^A =
\sigma^A/2$, with $\sigma^A$ the Pauli matrices), from which we can build all
the others. In this case we can write the link variable as $U_\mu(n)=
U_\mu^0(n) I_{2\times 2} + \mathrm{i}U_\mu^A(n)\sigma^A$ with
$\left[U_\mu^0(n)\right]^2 + \left[U_\mu^1(n)\right]^2 +
\left[U_\mu^2(n)\right]^2 + \left[U_\mu^3(n)\right]^2  = 1$. The partition
function becomes
\begin{equation}
\label{ZSU2a}
Z_{SU(2)} = \int\left[d U_\mu^0(n)d U_\mu^1(n)d U_\mu^2(n)d U_\mu^3(n)
\delta( \left[U_\mu^0(n)\right]^2 + \left[U_\mu^1(n)\right]^2 +
\left[U_\mu^2(n)\right]^2 + \left[U_\mu^3(n)\right]^2-1)
\right] e^{-S[U]}
\end{equation}
We insert now the expression
$$
1 = 
\prod_{\alpha=0}^3\int_{-\infty}^\infty dV_l^\alpha(n)\delta(V_l^\alpha(n)-U_\mu^\alpha(n))
$$
and use the integral representation of the delta functions to write the
partition function as 
\begin{equation}
\label{ZSU2b}
\begin{array}{l}
\displaystyle
Z_{SU(2)} = 
\int \left[d V_l^0(n)d V_l^1(n)d V_l^2(n)d V_l^3(n)
\frac{d\alpha_l^0}{2\pi}\frac{d\alpha_l^1}{2\pi}\frac{d\alpha_l^2}{2\pi}\frac{d\alpha_l^3}{2\pi}
\right]\\
\displaystyle
\hskip3truecm
 e^{-S[V]+\mathrm{i}\sum_l\left( \alpha_l^0V_l^0 + \alpha_l^1V_l^1 +
  \alpha_l^2V_l^2 + \alpha_l^3V_l^3\right) - \sum_l
  w(\alpha_l^0,\alpha_l^1,\alpha_l^2,\alpha_l^3)} 
\end{array}
\end{equation}
with 
\begin{equation}
\label{wSU2}
\begin{array}{l}
\displaystyle
e^{-w(\alpha_l^0,\alpha_l^1,\alpha_l^2,\alpha_l^3)}\equiv 
\int
\left[d U_\mu^0(n)d U_\mu^1(n)d U_\mu^2(n)d U_\mu^3(n)
\delta( \left[U_\mu^0(n)\right]^2 + \left[U_\mu^1(n)\right]^2 +
\left[U_\mu^2(n)\right]^2 + \left[U_\mu^3(n)\right]^2-1)
\right] \\
\displaystyle
\hskip3truecm
e^{-\mathrm{i}(\alpha_l^0 U_\mu^0(n)+\alpha_l^1 U_\mu^1(n)+\alpha_l^2
  U_\mu^2(n)+\alpha_l^3 U_\mu^3(n))}
\end{array}
\end{equation}
Since $\alpha_l^0 V_l^0 + \alpha_l^1 V_l^1 + \alpha_l^2 V_l^2 + \alpha_l^3
V_l^3= (1/2)\mathrm{Tr}\left[\alpha_l^\dagger\cdot V_l\right]$ and 
$w(\alpha_l^0, \alpha_l^1, \alpha_l^2, \alpha_l^3) =
w\left(\mathrm{Tr}\left[\alpha_l^\dagger\cdot\alpha_l\right]\right)$, we
confirm that the action is invariant under local $SU(2)$ transformations, $u_l$, that
act as $V_l\to u_l^\dagger V_l u_l$ and $\alpha_l\to u_l^\dagger \alpha_l
u_l$. The measure is similarly invariant and, therefore, the partition function
as well.  Once more we note that it is the mixing term,
$\mathrm{Tr}\left[\alpha^\dagger V\right]$ that breaks the $SU(2)\times SU(2)$
symmetry of the two other terms to $SU(2)$. 

In both cases we have ended up with field theories, whose fields live
on the links of the lattice. 
These fields have ``angular'' parts that take
values on the group manifold, $S^1$ for $U(1)$, $S^3$ for $SU(2)$ and radial
parts that are not ``frozen''--and that take values on one--dimensional
manifolds (this is related to the fact that both groups are of rank 1). They
may lead to more efficient numerical algorithms for computing the correlation
functions. Deferring this project, we shall try to evaluate the partition
functions by a saddle--point approach. The vacua we will focus on are uniform
field configurations, $V_l=V$ and $\alpha_l=\alpha$. For such configurations
all terms (in particular the Wilson action) now factorize over the links. The
action is  invariant only under {\em global} transformations on the
group manifolds. For the $U(1)$ case this means that one can generate all
saddle points from the configuration $\{(V^R, V^I), (\alpha^R,
\alpha^I)\}=\{(V,0),(\alpha,0)\}$ by acting with an arbitrary rotation matrix, 
$\mathrm{diag}(e^{\mathrm{i}\theta},e^{-\mathrm{i}\theta})$. If the solution 
we find by this {\em Ansatz}  is $V=0,\alpha=0$, then all rotations leave this
invariant: the global $U(1)$ symmetry is realized in the Wigner mode and the 
photon is ``confined''. If $V\neq 0, \alpha\neq 0$, then the global 
$U(1)$ symmetry is realized in the Nambu--Goldstone mode (it is spontaneously
broken, which is possible for a global symmetry) and the corresponding 
Goldstone boson is the photon. That it is massless is expressed by the fact
that a rotation of angle $\theta$ along the group manifold doesn't cost any
energy (cf. ref.~\cite{coleman_erice} for a similar discussion in the continuum).

For the $SU(2)$ case the calculation is
more interesting, since the variables are matrices. A uniform configuration 
in the fundamental representation, for example,  will be given by the matrix
$V\equiv V^0 I_{2\times 2} + \mathrm{i} \bm{V}\cdot\bm{\sigma}$
 and similarly for the matrix $\alpha\equiv \alpha^0 I_{2\times 2} + \mathrm{i}
\bm{\alpha}\cdot\bm{\sigma}$ (i.e. with constant coefficients).
  The different saddle points are related by the
action of an $SU(2)$ valued matrix $u$ in the fundamental representation, 
$u\equiv u_0 I_{2\times 2}+\mathrm{i}\bm{u}\cdot\bm{\sigma}$ with $u_0^2 +
||\bm{u}||^2=1$:  $V\to u^\dagger V u$ and $\alpha\to u^\dagger \alpha u$.  
It would seem that  we can't get {\em all} possible saddle points,
corresponding to uniform configurations, by such an action, if we restrict
ourselves to $V = V^0 I_{2\times 2}$ and $\alpha = \alpha^0 I_{2\times 2}$,
since $u^\dagger V u$ or $u^\dagger \alpha u$ don't move on a three-sphere, as
$u$ moves on the unit three-sphere: they stay put at the center  
(if $V^0=0,\alpha^0=0$) or at the ``North Pole'' (if $V^0\neq 0, \alpha^0\neq
0$) of the respective three-spheres. 
 We seem to need a richer set. Such a set is defined by including the 
element of the Cartan subalgebra, $\sigma_3$: 
$u^\dagger (V^0I_{2\times 2}+\mathrm{i}V^3\sigma_3)u$
does cover  a three-sphere of radius 
$\left(\left[V^0\right]^2+\left[V^3\right]^2\right)^{1/2}$, and, similarly, 
$u^\dagger(\alpha^0 I_{2\times 2}+\mathrm{i}\alpha^3\sigma_3)u$ covers a
three-sphere of radius 
$\left(\left[\alpha^0\right]^2+\left[\alpha^3\right]^2\right)^{1/2}$
as $u$ covers the three-sphere of unit radius. In other words, given any 
$SU(2)$ matrix, $V\equiv V^0 I_{2\times 2}  +
\mathrm{i}\bm{V}\cdot\bm{\sigma}$, there exists another $SU(2)$ matrix, 
$u\equiv u_0 I_{2\times 2} + \mathrm{i}\bm{u}\cdot{\bm\sigma}$, such that 
$u^\dagger V u = \widetilde{V}^0 I_{2\times 2} +
\mathrm{i}\widetilde{V}^3\sigma_3$. 
In this case the effective action for the mean field approximation of the
$SU(2)$ theory reads 
\begin{equation}
\label{SU2effact}
S_{SU(2)}[V,\alpha] = S\left[\left[\widetilde{V}^0\right]^2 + \left[\widetilde{V}^3\right]^2\right] +
\widetilde{\alpha}^0 \widetilde{V}^0 + \widetilde{\alpha}^3 \widetilde{V}^3 - 
w_{SU(2)}\left(\left[\widetilde{\alpha}^0\right]^2 + \left[\widetilde{\alpha}^3\right]^2\right)  
\end{equation}
and we notice something very interesting: it is invariant under $U(1)$
transformations! The {\em only} difference between this expression and the
corresponding one for the $U(1)$ theory (or any rank 1 group for that matter)
resides in the function $w(\cdot)$. So we  can obtain all {\em
  its} saddle points, by applying a $U(1)$ rotation to the configuration 
$(\widetilde{V}^0, \widetilde{V}^3=0,\widetilde{\alpha}^0, 
\widetilde{\alpha}^3 = 0)$. This reduces  our problem to that of 
finding the extrema of the action
\begin{equation}
\label{SU2effact1}
S_{SU(2)}[\widetilde{V}^0, 0, \widetilde{\alpha}^0,0] = 
S[\left[\widetilde{V}^0\right]^2] + \widetilde{\alpha}^0 \widetilde{V}^0 - w_{SU(2)}(\widetilde{\alpha}^0)
\end{equation}
We shall now show that we can obtain all saddle points, that are uniform
across the lattice, for the $SU(2)$ theory, from saddle points of the 
$U(1)$ theory,provided we  transform the couplings appropriately:
\begin{equation}
\beta_{SU(2)} = f(\beta_{U(1)}) 
\end{equation}
To establish this relation we argue as follows: 
The problem boils down to the study
of the family of functions of two variables, $v$ and $\alpha$, with $\beta$ a
parameter:
\begin{equation}
\label{effactionredux}
S[v,\alpha;\beta]\equiv \beta s(v)-\alpha v + w(\alpha)
\end{equation}
Its extrema are solutions of the equations 
\begin{equation}
\label{sols}
\begin{array}{ccc}
\displaystyle v = w'(\alpha) & \displaystyle \mathrm{and} &
\displaystyle \alpha = \beta s'(v)
\end{array}
\end{equation}
These two equations may be reduced to one,
\begin{equation}
\label{master_saddle}
\beta s'\left(w'(\alpha)\right) = \alpha
\end{equation}
The function $s(v)=1-v^4$, but the properties that we will really need are 
that $s'(0)=0$ (and only there).

We are interested in how the solutions of this equation behave as we vary the
function $w(\cdot)$, within the family of monotonic, differentiable,
functions, that vanish at the origin.
 We note that the value $\alpha=0$ is always a solution. We are
looking for non-trivial solutions and look to establish the equivalence
\begin{equation}
\label{SU2/U1equiv}
\beta_1s'\left(w_1'(\alpha_1)\right) = \alpha_1\Leftrightarrow
\beta_2s'\left(w_2'(\alpha_2)\right) = \alpha_2
\end{equation} 
This is, indeed, possible: Given the solution, $\alpha_1\neq 0$,
 of the first equation (that depends on $\beta_1$, of course), 
we can find $\alpha_2(\beta_1)\equiv w_2^{-1}\left(w_1(\alpha_1(\beta_1))\right)$. 
The second equation then allows us to obtain 
\begin{equation}
\label{SU2/U1beta}
\beta_2 = \frac{\alpha_2(\beta_1)}{s'(w_2'(\alpha_2(\beta_1)))}\equiv f(\beta_1)
\end{equation} 
 Therefore, if, for $\beta=\beta_1$ the first theory was in
the Coulomb phase, for $\beta=\beta_2$ the second theory will be as well: we
have an equivalence, not a duality. We remark that the definition of
$\alpha_2$ is not unique: The only requirement is that it allow us to find
$\beta_2$. This freedom is, indeed, an expression of the universality that, if
we have the solution, $\alpha_1(\beta_1)$, $v_1 = w_1'(\alpha_1(\beta_1))$, we
have the solution, in the mean field approximation for any rank 1, compact
gauge group.  
Furthermore, we realize that the assumption that the uniform field
configurations may be taken proportional to the identity is not an
additional approximation at all, but a consequence of the mean field
approach: had we not chosen such a configuration, we would have come across
it, when generating all the saddle points. If we could not have done so, we 
would have made a mistake.

The saddle point, $V^0=0,\alpha^0=0$ realizes the global $SU(2)$ symmetry in
the Wigner mode: if we act upon it by any $U(1)$ transformation, we still get
zero and the action of any $SU(2)$ element $u$ in the corresponding
representation still gives zero. The three gauge fields of $SU(2)$ are
confined. If we find a saddle point with $V^0\neq 0, \alpha^0\neq 0$ then
the global $SU(2)$ symmetry is spontaneously broken. We would expect three
Goldstone bosons, corresponding to the angular directions along $S^3$. 
If we act on the configuration $(V^0,V^3=0), (\alpha^0,\alpha^3=0)$ by a 
$U(1)$ rotation matrix, we will obtain the field along $S^1$, i.e. in the 
Cartan subalgebra, 
$V = V^0\left(\cos\theta I_{2\times 2}-\mathrm{i}\sigma_3\sin\theta\right),
\alpha = \alpha^0\left(\cos\theta I_{2\times
  2}-\mathrm{i}\sigma_3\sin\theta\right)$. 

Acting on {\em this}
 configuration with  some element $u^\dagger$ on the left and $u$ on the right
 we can find the coefficient functions of
 $\sigma^\pm\equiv\sigma_1\pm\mathrm{i}\sigma_2$ and $\sigma_3$ for the
 ``propagating'' modes, $W^\pm, W^3$ (and similarly for the constraints,
 generated by the transforms of the $\alpha$'s) 
\begin{equation}
\label{Wpm}
\begin{array}{l}
\displaystyle 
W^+ = \mathrm{Tr}\left[u^\dagger V u\sigma^+\right]\\
\displaystyle
W^- = \mathrm{Tr}\left[u^\dagger V u\sigma^-\right]\\
\displaystyle 
W^3 = \mathrm{Tr}\left[u^\dagger V u\sigma_3\right]\\
\end{array}
\end{equation}
When we insert these exressions in the action, the $u^\dagger$ and $u$ cancel
out, since the action is invariant under $SU(2)$ transformtions.
So, despite appearences, this configuration is gauge equivalent to a 
$U(1)$ gauge configuration, i.e. a ``photon'', since $u^\dagger,u$ are 
globally defined. 

 The fundamental reason is, of course, that the gauge field is taken
constant over the lattice-it is this simplification, that allows us to use a
global $SU(2)$ gauge transformation to rotate {\em all} links to a constant 
Cartan matrix, that, then may be rotated, by a global $U(1)$ transformation,
(that, of course, also is embeddable in $SU(2)$) to the identity. And these
transformations rely on the fact that the fields in question take values in
the group and not the algebra. This is  the difference
with the continuum, where a constant non-abelian gauge potential
configuration can give rise to a field strength that is not gauge equivalent
to an Abelian one--a manifestation of the ``Wu--Yang
ambiguity''~\cite{tudron}.  The reason
of the difference is that the gauge fields in the continuum take values in the
algebra, which is non-compact, whereas, on the lattice, they take values in
the group, which is compact, for the groups we are studying. This is
expressed by the functions $w_{SU(2)}$ and $w_{U(1)}$, which exist for the
(compact) groups but not the (non-compact) algebras. The correspondance
between the $U(1)$ and $SU(2)$ actions isn't possible, if these functions
don't exist, or are not invertible. Of course these conditions aren't
sufficient: the correspondance cannot (and will not) hold if higher
dimensional, non-uniform representations of $SU(2)$ are considered. 

Let us now see what happens when anisotropic couplings are
introduced~\cite{fu_nielsen,nicolis,fu_nielsenYM, knechtlietal,petrov}. 

Let us assume, for the moment, that there is one extra dimension. This means
that there don't exist plaquettes with only $V'$ links and that the mixing
term, $S_\mathrm{mix}[V,V']$ contains plaquettes with two $V$ and two $V'$
links only:
\begin{equation}
\label{mixing_term}
S_\mathrm{mix}[V', V] =
\beta'\sum_{links}\left(1-\mathrm{Re}\left(
\mathrm{Tr}\left[V'_{l_1}V_{l_2}{V'_{l_3}}^\dagger {V_{l_4}}^\dagger\right]\right)
\right)
\end{equation}
where the links $l_1$ and $l_3$ point in the extra dimension.

Absent the mixing term, the action is invariant under local $G\times G$
transformations ($U(1)\times U(1)$ or $SU(2)\times SU(2)$ for the examples we
are considering, but this holds for any gauge group, of course). We notice
that the $V'$ field does not have a plaquette term (which would be possible
if there were at least two extra dimensions). 

The mixing term is invariant only under 
{\em global} $G\times G$ transformations, even
for non-uniform field configurations:it explicitly breaks the {\em gauge}
symmetry (since it doesn't contain an oriented loop of gauge fields,
 that transform under the same representation of the group) so, generically, 
the theory can't describe the propagation of 
gauge particles. This distinguishes it from the ``kinetic
mixing''~\cite{kinetic_mixing} terms. 
 
If $\beta = \beta'$, however, then the action 
is invariant under local transformations of the diagonal sugbroup, $G$ of
$G\times G$ that can, then,  be consistently gauged. 
If $\beta\neq\beta'$ this isn't possible. 

If, however, 
 we can tune the couplings $\beta$ and $\beta'$ so that the mixing term
vanishes, or becomes a global constant 
(as such, in the mean field approximation; inside correlation
functions beyond it, i.e. by imposing the appropriate 
Ward--Takahashi identities), then the global symmetry can be promoted to a 
gauge symmetry, even if $\beta\neq \beta'$. 

When we consider uniform configurations where $V_l = V, \alpha_l = \alpha$
and $V'_l = V', \alpha'_l = \alpha'$ we are respecting the full symmetry
group.

In what follows, we discuss in what ways this can be achieved, within the
classes of solutions to the saddle point equations and how corrections might 
affect this.

The saddle point equations for uniform configurations have three classes of
solutions: 

(a) $(V = 0, \alpha = 0), (V' = 0, \alpha' = 0)$. In this phase the symmetry 
under $G\times G$ is realized in the Wigner mode. The mixed term doesn't
vanish, but becomes an irrelevant constant. In the mean field approximation
the situation seems trivial--the true test is, whether the corrections to the
mean field appoximation can satisfy 
$\mathrm{Re}\left[\mathrm{Tr}\left(
  {V_{l_1}}{V'_{l_2}}{V'_{l_3}}^\dagger{V_{l_1}}^\dagger\right)\right] = 1$
while giving rise to an area law for the Wilson loops, which, under this
condition, become well-defined. 
In fact they could satisfy another condition, namely, $V' V =
z V V'$, with $z$ in the center of the
group--and constant.  This imposes additional 
conditions on the couplings, $\beta$ and $\beta'$. 
The confining phase thus makes sense only on a given layer, 
since there isn't any symmetry ``protecting'' the Wilson loop, that would
 ``stick out'' in the extra dimension. So we have a 
layered, confining, phase. This is what one would expect for a Yang--Mills
theory. If $D=4$ this will hold~\cite{creutz} for $\beta = \beta'$. For $D>4$ 
this will hold only in the presence of a cutoff. 

If these additional conditions aren't, or cannot be,
satisfied, then the symmetry remains global and 
the theory describes fields on links, invariant under a global symmetry, whose
continuum limit must still be established. In more than four dimensions it
most likely would be a free field theory. 

(b) $(V\neq 0, \alpha\neq 0), (V'\neq 0, \alpha'\neq 0)$. In this phase the
symmetry under $G\times G$ is spontaneously broken. However we still have
difficulties in describing this as the Coulomb phase of a gauge theory, since 
the mixing term doesn't seem to allow us to gauge the symmetry. So we may only
speak of a ``bulk scalar phase'' in general.

If, once more, 
 $\mathrm{Re}\left[\mathrm{Tr}\left(V' V V'^\dagger V^\dagger\right)\right] =
1$, then the mixing term vanishes,
the symmetry can be promoted to a local symmetry and we are in a 
{\em bona fide}  Coulomb phase, since, now, we {\em can} gauge the symmetry. 

If $V' V = z V V'$, with $z$ a constant element, in the
center of the group, the mixing term also allows us to obtain a Coulomb
phase. The difference with the finite temperature case is that the 
extra dimension is assumed to be space-like (upon Wick rotation back to 
Minkowski). This Coulomb phase is realized in layers, since the $V'$ links
aren't dynamical for two reasons: there isn't any plaquette term for them and
they are constrained by the vanishing of the mixing term. The {\em full}
 gauge symmetry is realized on the layer in the Nambu--Goldstone mode, since 
the group manifold, spanned by the angular degrees of freedom, is
unaffected. Breaking $SU(2)$ to $U(1)$, for instance, would entail the
impossibility of recovering solutions, where $u^\dagger$ and $u$ lived on
$S^3$, but only on an $S^1$ submanifold. This is not the case. It could,
however, be realized if $V' V = z V V'$ holds  for $z$ not in the center of
$SU(2)$, but in that of $U(1)$, i.e. for $z=e^{\mathrm{i}\Phi}$ with $\Phi$ 
a real number and not simply $\pm I$.  

 One is tempted to call this situation ``flux compactification''
in a field theory setting and it will be interesting to investigate its 
properties in detail.

Even in the presence of a plaquette term for the $V'$ (i.e. with at least two
extra dimensions) the mixing term must vanish for the gauge symmetry to be
realized, since it breaks gauge symmetry by itself. 
We do not and cannot have a ``bulk Coulomb phase'' in this case. 
Therefore, for any number of extra dimensions, 
the Coulomb phase is realized on layers, which can, therefore, be defined 
by the vanishing of the mixing term. Within mean field theory this situation 
is gauge equivalent to that of an abelian configuration on any given layer. 
 
It will be interesting to see what corrections to mean field theory, that are
sensitive to the group structure, and/or numerical simulations can tell us 
about this situation (a first attempt has been carried out in
ref.~\cite{kurkela_pdf}). 

Scenaria similar to this discussion have appeared  within the context of 
``deconstruction'' models (cf.~\cite{deconstruct}) and in various proposals
for breaking gauge symmetry by Wilson lines~\cite{hosotani};
the difference with the discussion here is that the anisotropic lattice action
provides a dynamical basis for the deconstruction scenario and shows how a 
UV completion could be quantitatively studied.  It also shows how the flux,
required by the Wilson lines is not {\em ad hoc}, but  a requirement for gauge
symmetry to be realized at all. In both cases it provides a lattice framework 
for studying it quantitatively, beyond perturbation theory.

(c) $(V\neq 0, \alpha\neq 0, V'=0, \alpha'=0)$. This is what is, usually,
 called the layered phase. In this case the mixing term becomes an irrelevant
 constant in the mean field approximation, since $V'=0$ in the layered phase
 and the gauge symmetry is realized on layers, in the Nambu--Goldstone mode. 

In summary, the gauge symmetry, if it is realized at all, is realized in
layers. The fields that live on the links, that point in the extra dimensions, 
 decouple for this to occur. 

These observations allow us to explain the results of
ref.~\cite{fu_nielsenYM,knechtlietal,petrov} where the anisotropy, originally 
introduced for the compact $U(1)$ case~\cite{fu_nielsen}, was studied, 
under different approximation schemes (among them the mean field approximation
and corrections thereof) for the case of $SU(2)$ and $SU(3)$ gauge
groups and a layered, Coulomb, phase was also found in five dimensions. 
This sounds,indeed, very surprising, since 
four--dimensional Yang--Mills theories don't have a Coulomb
phase at zero temperature~\cite{creutz}: they go from confinement at 
strong coupling to asymptotic freedom at weak coupling. We propose that the
explanation lies in the fact that,
in the mean field approximation, these results are an inevitable consequence of 
the equivalence of the $U(1)$ theory, that {\em does} have a layered 
Coulomb phase, with the $SU(2)$ theory, that, in this approximation, is, 
indeed, ``color-blind'', since its saddle points are gauge equivalent to 
abelian ones. This remains true in the presence of corrections, 
that are proportional only to  the identity of the group, since these can not 
probe the non-abelian group structure. For the $SU(3)$ case one 
is tempted to conjecture that the equivalence with the $U(1)\times U(1)$ 
theory is responsible and that  the equivalence provides, indeed, a 
realization of the ``Abelian projection''proposed by 't Hooft~\cite{thooft} 
many years ago and since studied for understanding confinement~\cite{polikarpov} (this was also remarked upon by Hosotani~\cite{hosotani}).

Let us show explicitly how such a correspondance can be established,
in a way that does not rely on an expansion around a particular solution at
all, for the rank 1 case.  
We will simply assume the existence of a (in general) non-uniform saddle 
point, proportional to the identity for one group. Then we will show that this 
implies the existence of a (generically non-uniform) saddle point with the 
same physical properties, proportional to the identity, for the other gauge 
group. 

One way is to remark that the saddle point equations for the gauge group,
defined by the function $w_1(\cdot)$
\begin{equation}
\label{sp1}
\begin{array}{ccc}
\displaystyle
\beta_1\frac{\partial s}{\partial V_l} = \alpha_l & 
\displaystyle &
\displaystyle
V_l = \frac{d w_1}{d\alpha_l}
\end{array}
\end{equation}
imply the ``integrability conditions''
\begin{equation}
\label{incond1}
\beta_1 = \frac{\alpha_l}{\partial s/\partial V_l}\Leftrightarrow
\alpha_k\frac{\partial s}{\partial V_l} = 
\alpha_l\frac{\partial s}{\partial V_k} 
\end{equation}
Next, we note that we can perform a (link-by-link) change of variables, 
$\alpha_l = g(\eta_l)$, without changing the physics. 
 If, however, we choose
this as $\alpha_l = w_1^{-1}\left(w_2(\eta_l)\right)$, which is well defined,
since the groups are compact, then we are describing 
a saddle point configuration in the theory defined by gauge group function
$w_2(\cdot)$, since the information about the gauge group lies, by
construction, in this case, exclusively, in the function $w(\cdot)$ 
(this was already remarked in ref.~\cite{drouffe_zuber}), as may be seen,
explicitly through the relation 
$$
S = \beta s(\{V\}) - 
\sum_{\mathrm{links}}\left(
w_1^{-1}\left(w_2(\eta_l)\right) V_l - w_1\left(
w_1^{-1}\left(w_2(\eta_l)\right) \right) \right) = 
\beta s(\{V\}) - 
\sum_{\mathrm{links}}
\left(
w_1^{-1}\left(w_2(\eta_l)\right) V_l - w_2(\eta_l)
\right)
$$
Therefore, 
$$
\beta_2 = \frac{w_1^{-1}(w_2(\eta_l))}{\partial s/\partial V_l}
$$
where $\partial S_2/\partial\eta_l=0\Leftrightarrow V_l = d w_2(w_1)/dw_1$,
showing that we have parameterized the group $w_2$ by the coordinate $w_1$: we
have changed coordinates, not physics. 

It's useful to understand 
 where exactly did we rely on the assumption that the saddle points 
were proportional to the identity. The point where we relied on it was when we
defined $\alpha_l\equiv w_1^{-1}\left( w_2(\eta_l)\right)$. 
If, for instance, the  saddle point, $\{\widehat{V}_l,\eta_l\}$
 were a non-uniform configuration in the fundamental representation of $SU(2)$, then $w_1(\alpha) = w_2\left( \left[\eta_l^0\right]^2 + ||\bm{\eta}_l||^2\right)$.
 Therefore, the previous equation simply fixes the norm of the vector
 $(\eta^0,\bm{\eta})$ and is blind to the individual
 components (this would be the case also, if the configuration were
 non-uniform within the Cartan).

These calculations also suffice to show that, in the anisotropic case, a
non-uniform configuration, proportional to the identity of $SU(2)$ (or any
other rank 1 group for that matter) {\em in the layered phase} is, in fact,
equivalent to a non-uniform $U(1)$ configuration, within the layer. (This
statement is very easy to understand in the continuum, where the fact that the
configurations are proportional to the identity means that the commutator term 
in the field strength vanishes identically~\cite{tudron}. On the lattice we
needed to take the group functions into account.)

In conclusion we have used an exact transcription of a lattice gauge theory,
which is interesting in its own right, to
obtain an equivalence between the dynamics of all pure gauge 
theories that are invariant under compact groups of rank 1 in the mean field
approximation, including a certain class of corrections to it, namely those
that remain proportional to the identity element of the groups. 

We have also clarified the phase structure, in the presence of anisotropy, 
and shown that the transition lines, that separate the layer phase from the 
others, are the only places where a gauge theory can, eventually, be defined
at all.  This shows explicitly that the mixing term decouples, when this is 
the case. We can define confining, as well as Coulomb, phases on these layers 
in an intrinsic way and it will be interesting to establish that the
conditions on the Wilson loops, that have been conjectured here 
(their mean field avatars have been established), do indeed hold and study 
``kinetic mixing'' actions~\cite{kinetic_mixing} on the lattice in this fashion.

Recently~\cite{nicolis10}  we have presented the analytical calculation that 
shows how the anisotropy leads to a second order phase transition between the 
layer phase and the bulk ``scalar'' phase, 
within the mean field approximation. An implication of
this calculation is that, indeed, 
along the layered to ``bulk scalar'' transition line, the {\em gauge} symmetry
 is recovered. 

In parallel with the computation of the corrections to the mean field
approximation, numerical simulations of the Ward--Takahashi identities are
essential towards clarifying the realization of the putative symmetries,
especially when taking into account coupling to matter, that has been left out
here. 

It might also be interesting to study the modulated phase in five--dimensional
gauge theories~\cite{ooguri} in this context.
  
For the case of higher rank groups (such as $SU(3)$ which is of rank 2) this
formulation also has certain advantages, namely it allows us to 
quantitatively study  the obstructions towards  realizing the  
abelian projection program. The computations are considerably more involved 
and will be presented elsewhere.

{\bf Acknowledgments:} It is a pleasure to acknowledge 
discussions with Ph. de Forcrand, E. G. Floratos and J. Iliopoulos.

\end{document}